# Anisotropic Fermi Surface and Quantum Limit Transport in High Mobility 3D Dirac Semimetal $Cd_3As_2$

Yanfei Zhao[1,2], Haiwen Liu[1,2], Chenglong Zhang[1,2], Huichao Wang[1,2], Junfeng Wang[3], Ziquan Lin[3], Ying Xing[1,2], Hong Lu[1,2], Jun Liu[4], Yong Wang[4], Shuang Jia[1,2,‡], X. C. Xie[1,2,†] and Jian Wang[1,2,*]

[1]*International Center for Quantum Materials, School of Physics, Peking University, Beijing 100871, China*

[2]*Collaborative Innovation Center of Quantum Matter, Beijing 100871, China*

[3]*Wuhan National High Magnetic Field Center, Huazhong University of Science and Technology, Wuhan 430074, China*

[4]*Center of Electron Microscopy, State Key Laboratory of Silicon Materials, Department of Materials Science and Engineering, Zhejiang University, Hangzhou, 310027, China*

Correspondence authors
[*] jianwangphysics@pku.edu.cn (J.W.)
[†] xcxie@pku.edu.cn (X.C.X.)
[‡] gwljiashuang@pku.edu.cn (S.J.).

**ABSTRACT**

The three-dimensional (3D) topological Dirac semimetal is a new topological phase of matter, viewed as the 3D analogy of graphene with a linear dispersion in the 3D momentum space. Here, we report the angular dependent magnetotransport in $Cd_3As_2$ single crystal and clearly show how the Fermi surface evolves when tilting the magnetic field. Remarkably, when the magnetic field lies in [112] and [44$\bar{1}$] axis, only single oscillation period features present, however, the system shows double period oscillations when the field is applied along [1$\bar{1}$0] direction. Moreover, tilting the magnetic field at certain direction also gives double period oscillations. We attribute the anomalous oscillation behavior to the sophisticated geometry of Fermi surface and illustrate a complete 3D Fermi surfaces with two nested anisotropic ellipsoids around the Dirac point. Additionally, a sub-millimeter mean free path at 6 K is observed in $Cd_3As_2$ crystal, indicating a large ballistic transport region in this material. Tracking the magnetoresistance oscillations to 60 T, we reach the quantum limit (n = 1 Landau Level) at about 43 T. These results improve the knowledge of the Dirac semimetal material $Cd_3As_2$, and also pave the way for proposing new electronic applications based on 3D Dirac materials.

## I. INTRODUCTION

The three-dimensional (3D) topological Dirac semimetal, a new type of topological materials with a pair of nondegenerate 3D massless Weyl fermions becomes a rapid growing field of research in condensed matter physics [1-13]. In analogy to two dimensional Dirac points observed in graphene [14] and topological insulators [15-19], 3D Dirac semimetal possesses bulk Dirac fermions with linear dispersion along three momentum directions [20].

Motivated by theoretical predications [1,2], to date, $Na_3Bi$ [3] and $Cd_3As_2$ [4,5,10,21] have been identified to be 3D topological Dirac semimetal by angle resolved photoemission spectroscopy (ARPES). More recently, scanning tunneling microscopy (STM) experiments have revealed quasiparticle interference and the extended Dirac-like dispersion in $Cd_3As_2$ [6]. This unusual band dispersion also makes the electrons around the Fermi surface behave many unusual transport phenomena such as strong quantum oscillations [8], ultrahigh mobility [9] and large magnetoresistance [9,12]. However, most magnetotransport measurements have been limited to study the quantum oscillations in one direction or by rotating the magnetic field only in certain plane. The scarcity of the complete 3D Fermi surface analysis hinders an in-depth understanding of the physical properties in $Cd_3As_2$ single crystal. Therefore, it is highly desirable to demonstrate the transport property study at different magnetic field direction as well as the angular dependent magneotransport to reveal the complicated 3D Fermi surface in $Cd_3As_2$ system. Further, the high magnetic field transport measurements to reveal the physics in quantum limit of $Cd_3As_2$ are also highly pursued.

Here, we present a systematic study of the magnetotransport in $Cd_3As_2$ single crystal and firstly extend our study to the angular dependence of magnetoresistance in three independent directions as well as the high magnetic field experiments (up to 60 T). The SdH oscillations signified a previously unknown Fermi surface with two nested ellipsoids, leading to a good understanding on its 3D Dirac nature and also providing a platform to explore exotic physical phenomena.

## II. RESULTS
### A. Sample structure

Single crystals of $Cd_3As_2$ are synthesized from a Cd-rich melt with the stoichiometry $Cd_8As_3$ in the evacuated quartz ampoule [22]. The $Cd_3As_2$ single crystal is examined by a FEI TITAN Cs-corrected cross sectional scanning transmission electron microscopy (STEM) operating at 200kV. Fig. 1(a) shows the atomic layer by layer high angle annular dark field scanning transmission electron microscopy (HAADF-STEM) image which manifests a high quality single crystal nature of $Cd_3As_2$ sample. An optical image of the measured sample is shown in the inset of Fig. 1(d). The crystal is needle-like and grows preferentially along the $[1\bar{1}0]$ direction (the length direction). The width direction is along $[44\bar{1}]$ and the largest facet of the crystal is (112) plane. Standard four probe method is used to measure the $Cd_3As_2$ samples on its (112) plane. Two indium or silver paste current electrodes (I+ and I-) are pressed on both ends and across the entire width of the sample, so that the current can homogeneously go through the sample in the length direction [1$\bar{1}$0]. The other two indium or silver paste electrodes were pressed on the crystal as voltage probes. Angular dependence of magnetoresistance was measured by rotating the sample in (1$\bar{1}$0) plane and (112) plane, characterized by the angle θ and φ illustrated in Fig. 1(b), respectively. The transport measurements were carried out in a PPMS-16T system (Quantum Design) and pulsed high magnetic field at Wuhan National High Magnetic Field Center.

**B. Ultrahigh mobility in Cd$_3$As$_2$ single crystal**

More than ten samples have been studied. Data presented here are from three typical samples: Sample 1, 2 and 3. Fig. 1(c) shows the resistivity of sample 1 as a function of temperature (*T*). The resistivity decreases almost linearly when *T* decreases from 300 K to about 6 K and then tends to be saturated. It displays a perfect linear metallic property with a large residual resistivity ratio RRR = 2120 (the resistivity at room temperature over the resistivity at 6 K). As shown in the insert panel of Fig. 1(c), the resistivity ρ is quite low (about $11.6\,n\Omega\cdot cm$ at 6 K). Similar results have been reported before [9], suggesting the behavior is related to its long transport lifetime. More interestingly, when *T* is lower than 6 K, the resistivity oscillates near zero resistivity within the resolution of our measurement instrument (about $4.5\,n\Omega\cdot cm$), which could be understood in terms of the quantum ballistic transport. Estimated from the Shubnikov-de Haas (SdH) oscillations of sample 1, the carrier density is about $5.86\times 10^{18}\,cm^{-3}$ (details shown in Supplemental Material [23]). It is worth noting that the carrier density estimation from SdH might be two to ten times smaller than the carrier density in Hall measurement [9]. Therefore, the mobility is conservatively estimated to be $9.19\times 10^6 \sim 4.60\times 10^7$ cm$^2$/Vs and the mean free path $l_0$ is about 0.25 ~ 1.25 mm at 6 K, which are consistent with the previous transport study [9]. The high mobility and long mean free path may shed new light on realizing various applications on future functional devices. Additionally, the linear metallic ρ – T behavior is also observed in other samples, such as sample 2 shown in Fig. 1(d).

**C. 3D Anisotropic Fermi surface**
**1. Transport properties**

To investigate the 3D shapes of Fermi surfaces of Cd$_3$As$_2$, the analysis of SdH oscillations measured in three independent magnetic field directions, along [112], [44$\bar{1}$] and [1$\bar{1}$0] axes (B$_{[112]}$, B$_{[44\bar{1}]}$ and B$_{[1\bar{1}0]}$) are shown in Fig. 2. Fig. 2(a) to (c) displays the field dependencies of resistivity measured in three magnetic fields at different temperatures. Pronounced SdH oscillations are clearly visible around 4 T up to 50 K, and then diminished at higher temperatures for all field directions. After subtracting a fourth-order polynomial background, the oscillatory component Δρ in three field directions are plotted in Fig. 2(d) to (f), respectively. The frequency of the SdH oscillations was extracted from the fast Fourier transform (FFT) analysis. Only one fundamental frequency is observed in both $B_{[112]}$ and $B_{[44\bar{1}]}$ directions with the period $B_{F[112]}$ = 54.30 T and $B_{F[44\bar{1}]}$ = 43.44 T, indicating a relatively simple but anisotropic Fermi surface in the two directions and the Fermi surface in $B_{[44\bar{1}]}$ direction is smaller than that measured in $B_{[112]}$ direction. Interestingly, Fig. 2(f) illustrates a different set of SdH oscillations when the magnetic field is applied parallel to $[1\bar{1}0]$ axis. Compared with the magnetoresistance measured along [112] and $[44\bar{1}]$ direction, the SdH oscillations in $B_{[1\bar{1}0]}$ direction are obviously weaker and nonperiodic. At around 8 T, the oscillation seems to be suppressed and then appears again at about 10 T. The inset of Fig. 2(f) shows the FFT of SdH oscillations in $B_{[1\bar{1}0]}$ direction, two well-defined peaks appear at $F_1$ = 43.44 T and $F_2$ = 54.30 T. According to the Onsager relation, the frequency of the magnetoresistance oscillation is related to the external cross-sectional area $A_k$ of the Fermi surface in the momentum space: $F = \frac{\hbar}{2\pi e} A_k$, thus, gives two cross-sectional areas of Fermi surface perpendicular to the field $A_{F1} = 4.17 \times 10^{-3}$ Å$^{-2}$ and $A_{F2} = 5.18 \times 10^{-3}$ Å$^{-2}$, indicating a complicated Fermi surface.

In order to verify and analyze the SdH oscillations, we fit the entire oscillatory component with the standard Lifshitz−Kosevich (LK) theory for a 3D system [24-26]

$$\Delta\rho \propto \frac{\lambda}{\sinh\lambda} e^{-\lambda_D} \cos 2\pi \left[ \frac{F}{B} + \gamma - \delta \right] \quad (1)$$

with $\lambda = 2\pi^2 k_B T m^* / \hbar eB$ and $\lambda_D = 2\pi^2 k_B T_D m^* / \hbar eB$, where m$^*$ is the cyclotron effective mass of the carriers and $T_D$ is the Dingle temperature. γ is the phase related to the berry phase while δ is a phase shift determined by the dimensionality, taking the value δ = ±1/8 for 3D system [25,27]. Considering the two frequencies obtained in $B_{[1\bar{1}0]}$ direction, we use two independent parameters $F_1$ and $F_2$ to fit the oscillation. As shown by the red solid line in Fig. 2(f), fitting to the LK theory yields two frequencies $F_1$ = 44.37 T and $F_2$ = 52.85 T. The two frequencies almost consistent with the FFT results $F_1$ = 43.44 T and $F_2$ = 54.30 T, confirming the complex physics in this direction. On the other hand, the LK theory fitting to the oscillations in $B_{[112]}$ and $B_{[44\bar{1}]}$ directions with a single frequency are shown in the Supplemental Material [23]. The fitting to the oscillation in $B_{[112]}$ direction yields $F_{[112]}$ = 54.13 T, $T_D$ = 36.93 ± 1.5 K, and $\gamma-\delta$ = 0.3, respectively. For the $B_{[44\bar{1}]}$ case, the good fitting gives $F_{[44\bar{1]}}$ = 42.6 T, $T_D$ = 33.54 ± 2.5 K, and $\gamma-\delta$ = 0.2. Both of the fitting results are consistent with the frequency identified from FFT spectra ($F_{[112]}$ = 54.30 T and $F_{[44\bar{1}]}$ = 43.44 T). From the temperature dependence of the oscillation amplitude (Fig. 2(d) and 2(e)), we obtain the cyclotron effective mass $m^*_{[112]}$ = 0.043 $m_e$ and $m^*_{[44\bar{1}]}$ = 0.036 $m_e$ ($m_e$ is the free electron mass). Therefore, taking into account all three different directions, the observed periods point to a 3D Fermi surface is a simple anisotropic ellipse along both [112] and [44$\bar{1}$] direction, while, with complicated geometry along [1$\bar{1}$0] axis.

Moreover, to further understand the anisotropic 3D Fermi surface, the analysis of nontrivial Berry phase deduced from SdH oscillations is necessary. The Landau Level fan diagram of $B_{[112]}$ direction is plotted in Fig. 3(a). The maxima of $\rho$ are assigned to be the integer indices (solid circles) while the minima of $\rho$ are plotted by open circles in the diagram as half integer indices [25]. A linear extrapolation of the index plot gives the intercept value close to 0.3. Considering the study before [25], in a 3D system, the intercept of the index plot should be 0 ± 1/8 (+ for holes and − for electrons). In our system, the intercept is 0.3, deviating from ±1/8, indicating the Fermi surface is an anisotropic ellipsoid instead of spherical with perfect π Berry phase. Similarly, the LL fan diagram in $B_{[44\bar{1}]}$ direction is shown in Fig. 3(b). As discussed previously, the intercept value obtained is about 0.2, different from 0.3 yielded in $B_{[112]}$ direction, also consistent with the assumption of the anisotropic Fermi surface in Cd$_3$As$_2$ system. Fig. 3(c) displays the magnetoresistance behavior measured in pulsed high magnetic field normal to the (112) plane up to 54 T at various temperatures. The larger and more obvious SdH oscillations in high field provide an opportunity to study the further physics in the quantum limit. However, limited by the measurement resolution, the small oscillations in the lower field are hard to distinguish from noise. A similar plot based on the maxima and minima of $\rho$ versus the index n is shown in Fig. 3(d). The intercept shifts to 0.38 instead of 0.3 deduced from the relatively low magnetic field. It suggests that the high magnetic field is needed to fix the intercept more reliable. It is worth noting that we reach the quantum limit in sample 3 and more details are shown in section D.

To study the 3D Fermi surface of Cd$_3$As$_2$ systematically, the angular dependence of SdH oscillations are important to show the evolution of Fermi surface changing in different directions. Fig. 4(a) shows the angular dependent oscillations after removing the polynomial background for the transverse rotation (B⊥I) by varying magnetic field angle θ tilting from perpendicular field ($B_{[112]}$) to parallel field ($B_{[44\bar{1}]}$). θ is the angle between magnetic field and [112] axis. Fig. 4(b) shows the angular-dependent SdH oscillations for the transverse to longitudinal rotation in (112) plane (*I* was applied along [1$\bar{1}$0] and B was rotated from [44$\bar{1}$] towards [1$\bar{1}$0]). φ is the angle between magnetic field and [44$\bar{1}$] axis. As shown in Table I, the FFT analysis gives out when the magnetic field lies in [112] and [44$\bar{1}$] axis, only one period features present, however, the system shows double period oscillations when the field applied along [1$\bar{1}$0] axis. Moreover, when changing the magnetic field angle θ and φ, two period oscillations also present. We attribute these anomalous SdH oscillations to the sophisticated geometry of Fermi surface.

**2. Theoretical analysis**

From pervious theoretical and experimental results [2,4], the Fermi surface is two anisotropic ellipsoids around the Dirac point, with Fermi momentum $k_z$ much larger than $k_x$ and $k_y$. Moreover, when the Fermi energy goes beyond the Lifshitz saddle point, these two separate ellipsoids finally change into two nested ellipsoids. Therefore, the anomalous two-period SdH oscillations in $B_{[1\bar{1}0]}$ direction maybe essentially originate from these two nested ellipsoid Fermi surface, as shown in Fig. 5(c).

To further analysis the nested ellipsoid Fermi surface beyond the Lifshitz saddle point, we calculate the Fermi wave vectors at different directions. Based on the SdH period $S_{F[112]}$ = 54.30 T and $S_{F[44\bar{1}]}$ = 43.44 T for $B_{[112]}$ and $B_{[44\bar{1}]}$ direction, we obtained the mean momentum $k_{[112]}$= 0.041 Å$^{-1}$ and $k_{[44\bar{1}]}$= 0.036 Å$^{-1}$, implying an anisotropic rather than a spherical Fermi surface, which is qualitatively consistent with ARPES results [4]. Next, we analysis the anomalous SdH oscillations observed in the angular dependent measurements shown in Table I. This anomalous SdH oscillation can be attributed to the nested ellipsoid Fermi surface beyond the Lifshitz saddle point. Firstly, for the $B_{[112]}$ and $B_{[44\bar{1}]}$ direction, the maximum cross-section of Fermi surface does not pass through the nested region [Fig. 5(a) and (b)], thus, giving only single period. When the magnetic field is twisted from [112] to [44$\bar{1}$], the maximum cross-sections of Fermi surface penetrate into the nested region at certain angle range, and form two nested ellipses as shown in Fig. 5(d). Secondly, when magnetic field is twisted from [44$\bar{1}$] to [1$\bar{1}$0], the maximum cross-section of Fermi surface also form two nested ellipses when the direction of magnetic field is around [1$\bar{1}$0] direction [Fig. 5(c) and (d)]. The two nested ellipses lead to the anomalous two-period SdH oscillaitons in Table I. As shown in Fig. 5(d), the small dashed (red or blue) ellipse leads to the unchanged period $F_1 = 43.44$ T, while the overlapping region changes all the time, corresponding to the varied period $F_2$ shown in Table I. Therefore, the geometry of maximum cross-sections in Fermi surface can qualitatively interpret the anomalous SdH oscillations when tilting the magnetic field angle. (See the Supplemental Material for further analysis Ref. [23])

From the above analysis, we know that the Fermi energy is above the Lifshitz saddle point and the Fermi surface merged in certain momentum region. Thus, the topological correction from all two Dirac point contribute to the Landau level intercept value, which makes the intercept value deviates from the theoretical value of - 1/8 for electron type and results in different intercept values for different magnetic field orientations, which is consistent with the STM results [6]. In addition, comparing the Fermi surface with the previous reported Wigner-Seitz unit cell [2], we conclude that the Fermi surface in $Cd_3As_2$ is very large. Due to this large Fermi surface, the Umklapp relation [28] can be satisfied and the Umklapp electron-phonon scattering processes plays dominant role on the resistivity at low temperature in $Cd_3As_2$, which leads to $R \sim T \cdot N_{ph}$ with $N_{ph}$ denoting the number of phonons that satisfies the Umklapp relation [28]. Moreover, due to the ultra-large unit cell of $Cd_3As_2$, low energy optical phonon modes might exist in the system. Therefore, the Umklapp processes and optical phonon modes can lead to almost linear R-T relation down to very low temperature, and the observed linear R-T behavior in Fig. 1(c) and (d) deviates from the ordinary Bloch-Grüneisen Law with $R \sim T^5$ and the electron-electron interaction induced $R \sim T^2$ law [28].

**D. Quantum limit of $Cd_3As_2$ single crystal: SdH oscillations with Zeeman splitting and linear magnetoresistance in the quantum limit**

We extend the SdH measurements to 60 T to explore the quantum limit. According to the theory before [29], a quantum linear magnetoresistance would be expected to occur in the quantum limit in a gapless semiconductor with linear energy dispersion, where all the carriers occupy the lowest Landau Level. Fig. 6(a) displays the SdH oscillations of sample 3 in the $B_{[112]}$ direction measured in the pulsed high magnetic field at $T = 4.2$ K. At low field (B < 15 T), the magnetoresistance shows perfect SdH oscillations, while in the relative higher field (15 T < B < 43 T), it exhibits the SdH oscillations with Zeeman splitting. Careful investigation shows obvious Zeeman splitting is around n = 2 peak and n = 1.5 dip. The Landau level index demonstrates the system reaches the quantum limit at about 43 T. With increasing the magnetic field beyond the quantum limit (B > 43 T), the magnetoresistance presents a linear behavior which is consistent with the theoretical consideration [29]. After subtracting the polynominal background [the inset of Fig. 6(a)], the SdH oscillations with Zeeman splitting feature can be clearly observed. By analyzing the FFT of the oscillation, we yield a single frequency $F_{[112]} = 50.09$ T. Thus, comparing with sample 2, the relative small Fermi surface and the remarkable Zeeman splitting in sample 3 might be the reason for the appearance quantum limit at B = 43 T in sample 3. Further studies are needed to uncover the exact cause. Moreover, with considering the similar anisotropic Fermi surface as sample 2, the carrier density can be estimated as $n_{sdH} = 5.19 \times 10^{18}\ cm^{-3}$. The Landau Level index plot based on the positions of maxima and minima in ρ as a function of 1 / B is shown in Fig. 6(b), with the intercept is about 0.11. When approaching the quantum limit of n = 1, the maxima (or minima) points shows small deviations from the fitting line, which originates from the anomalous oscillation in 3D Dirac system around the Quantum limit [30].

## III. CONCLUSIONS

In summary, we firstly report pronounced SdH oscillations measured along three different directions in high quality $Cd_3As_2$ single crystals, and extend our study to the angular dependent magnetotransport and in the high magnetic field up to the quantum limit. By analyzing the SdH oscillations in different magnetic field orientations, we obtain a complete 3D Fermi surface with two nested ellipsoids. Furthermore, we present the changing of the angular-dependent oscillation periods are essentially due to the complicated nested 3D Fermi surface. In addition, the sub-millimeter scale mean free path and ballistic transport region as well as the quantum limit have been demonstrated in $Cd_3As_2$ single crystal. These results intensify the previous knowledge of the Dirac semimetal material $Cd_3As_2$, offer a better understanding of existing 3D Dirac semimetals, and reveal the potential of application in topological electronic devices.

**Acknowledgments**

Y.Z. and H.L contribute equally to this work. We acknowledge Chong Wang, Yuan Li, Tian Qian and Hong Ding for the help in Laue measurements and thank Liang Li and Zhengcai Xia for helpful discussions about the pulsed magnetic field measurements. This work was financially supported by National Basic Research Program of China (Grant Nos. 2013CB934600 and 2012CB921300), the National Natural Science Foundation of China (Nos. 11222434, 11174007), and the Research Fund for the Doctoral Program of Higher Education (RFDP) of China.

FIG. 1 (color online). Sample structure of $Cd_3As_2$ single crystal. (a) High angle annular dark field scanning transmission electron microscopy (HAADF-STEM) image of $Cd_3As_2$ single crystal. Scale bar represents 1 nm. (b) Schematic structure for the angular dependent magnetotransport measurements in $Cd_3As_2$ system. Sample is rotated in $(1\bar{1}0)$ plane and (112) plane, characterized by the angle θ and φ, respectively. The sample size is not to scale. (c) Resistivity of sample 1 as a function of temperature. The inset shows the resistivity falling to quite low value about $11.6\, n\Omega \cdot cm$ at 6 K and then reaching zero resistivity within instrumental resolution. (d) ρ (T) behavior of Sample 2. The inset shows an optical image of the measured $Cd_3As_2$ sample. Standard four probe method is used to measure the transport property.

FIG. 2 (color online). SdH oscillations for $Cd_3As_2$ single crystal in different magnetic field directions, along the [112], $[44\bar{1}]$ and $[1\bar{1}0]$ axes ($B_{[112]}$, $B_{[44\bar{1}]}$ and $B_{[1\bar{1}0]}$). (a)-(c) Magnetoresistivity measured in the perpendicular field $B_{[112]}$, the parallel fields $B_{[44\bar{1}]}$ and $B_{[1\bar{1}0]}$ at different temperatures, respectively. (d)-(e) The oscillatory component of Δρ extracted from ρ by subtracting a forth-polynomial background, as a function of 1/B at varies temperatures in $B_{[112]}$ and $B_{[44\bar{1}]}$ directions. (f) The LK theory fitting to the Δρ (B) curve measured at $T$ = 2 K in the parallel field $B_{[1\bar{1}0]}$ with two frequencies (red solid curve). Inset shows the FFT analysis with two frequencies in $B_{[1\bar{1}0]}$ direction.

FIG. 3 (color online). Landau level index plots for $Cd_3As_2$ single crystal in $B_{[112]}$ and $B_{[44\bar{1}]}$ directions. (a) Landau level index plot of the Fermi surface in the perpendicular field $B_{[112]}$. The maxima of Δρ are assigned to be the integer indices (solid circles) while the minima of Δρ are plotted by open circles as half integer indices. The intercept is close to 0.3. (b) Landau level plot for oscillations in Δρ measured at 2 K in the parallel field $B_{[44\bar{1}]}$. Solid line is a linear fitting to the data, giving the intercept about 0.2. (c) SdH oscillations measured in the pulsed high magnetic field in the perpendicular field $B_{[112]}$. (d) The index plot of 1/B versus n measured in pulsed high magnetic field. The intercept is about 0.38.

FIG. 4 (color online). Angular dependent of SdH oscillations at different magnetic field directions. (a) SdH oscillations in Δρ for various magnetic field angles from $B_{[112]}$ (θ = 0 deg, B in the [112] direction) to $B_{[44\bar{1}]}$ (θ = 90 deg, B in the $[44\bar{1}]$ direction). (b) SdH oscillations rotates from $B_{[44\bar{1}]}$ (φ = 0 deg, B in the $[44\bar{1}]$ direction) to $B_{[1\bar{1}0]}$ (φ = 90 deg, B in the $[1\bar{1}0]$ direction). Fig. 1(b) depicts the definition of θ and φ.

FIG. 5 (color online). Theoretical analysis of 3D Nested Anisotropic Fermi Surface in $Cd_3As_2$.

(a)-(c) show the largest cross section of Fermi surface versus the magnetic field orientation. The Fermi surface is two nested anisotropic ellipsoids. (d) shows the schematic plot of the Fermi surface for $B_{[1\bar{1}0]}$ direction with an overlapping region. The figures are not drawn to scale.

FIG. 6 (color online). High magnetic field transport measurement of Sample 3 up to 60 T. (a) Magnetoresistivity measured in the perpendicular field $B_{[112]}$ at different temperatures at $T$ = 4.2 K up to 60 T. Zeeman splitting features can be observed at higher magnetic field. Inset shows the SdH oscillations after subtract the polynominal background. (b) Landau level index plot. The n = 1 Landau Level is reached at B = 43 T. Quantum linear magetoresistance is observed at the field higher than 43 T.

**References**

[1] Z. J. Wang, Y. Sun, X.-Q. Chen, C. Franchini, G. Xu, H. M. Weng, X. Dai, and Z. Fang, *Dirac semimetal and topological phase transitions in $A_3Bi$ (A = Na, K, Rb)*, Phys. Rev. B **85**, 195320 (2012).

[2] Z. J. Wang, H. M. Weng, Q. S. Wu, X. Dai, and Z. Fang, *Three-dimensional Dirac semimetal and quantum transport in $Cd_3As_2$*, Phy. Rev. B **88**, 125427 (2013).

[3] Z. K. Liu, B. Zhou, Y. Zhang, Z. J. Wang, H. M. Weng, D. Prabhakaran, S.-K. Mo, Z. X. Shen, Z. Fang, X. Dai, Z. Hussain, and Y. L. Chen, *Discovery of a Three-Dimensional Topological Dirac Semimetal, $Na_3Bi$*, Science **343**, 864 (2014).

[4] Z. K. Liu, J. Jiang, B. Zhou, Z. J. Wang, Y. Zhang, H. M. Weng, D. Prabhakaran, S.-K. Mo, H. Peng, P. Dudin, T. Kim, M. Hoesch, Z. Fang, X. Dai, Z. X. Shen, D. L. Feng, Z. Hussain, and Y. L. Chen, *A stable three-dimensional topological Dirac semimetal $Cd_3As_2$*, Nat. Mater. **13**, 677 (2014).

[5] M. Neupane, S.-Y. Xu, R. Sankar, N. Alidoust, G. Bian, C. Liu, I. Belopolski, T.-R. Chang, H.-T. Jeng, H. Lin, A. Bansil, F. C. Chou, and M. Z. Hasan, *Observation of a three-dimensional topological Dirac semimetal phase in high-mobility $Cd_3As_2$*, Nat. Commun. **5**, 3786 (2014).

[6] S. Jeon, B. B. Zhou, A. Gyenis, B. E. Feldman, I. Kimchi, A. C. Potter, Q. D. Gibson, R. J. Cava, A. Vishwanath, and A. Yazdani, *Landau quantization and quasiparticle interference in the three-dimensional Dirac semimetal $Cd_3As_2$*, Nat. Mater. **13**, 851 (2014).

[7] Z. Zhu, and J. E. Hoffman, *Condensed-matter physics: Catching relativistic electrons*, Nature **513**, 319 (2014).

[8] L. P. He, X. C. Hong, J. K. Dong, J. Pan, Z. Zhang, J. Zhang, and S. Y. Li, *Quantum transport in the three-dimensional Dirac semimetal $Cd_3As_2$*, arXiv:1404.2557; Phys. Rev. Lett. **113**, 246402 (2014).

[9] L. Tian, Q. Gibson, M. N. Ali, M. Liu, R. J. Cava, and N. P. Ong, *Ultrahigh mobility and giant magnetoresistance in $Cd_3As_2$: protection from backscattering in a Dirac semimetal*, arXiv:1404.7794; Nat. Mater. doi: 10.1038/NMAT4143 (2014).

[10] S. Borisenko, Q. Gibson, D. Evtushinsky, V. Zabolotnyy, B. Büchner, and R. J. Cava, *Experimental Realization of a Three-Dimensional Dirac Semimetal*, Phys. Rev. Lett. **113**, 027603 (2014).

[11] Q. D. Gibson, L. M. Schoop, L. Muechler, L. S. Xie, M. Hirschberger, N. P. Ong, R. Car, and R. J. Cava, *3D Dirac semimetals: current materials, design principles and predictions of new materials*, arXiv:1411.0005.

[12] J. Y. Feng, Y. Pang. D. S. Wu, Z. J. Wang, H. M. Weng, J. Q. Li, Z. Fang, Y. G. Shi, and L. Lu, *Large linear magnetoresistance in Dirac semi-metal $Cd_3As_2$ with Fermi surfaces close to the Dirac points*, arXiv:1405.6611.

[13] S. Zhang, Q. Wu, L. Schoop, M. N. Ali, Y. G. Shi, N. Ni, Q. Gibson, V. Sidorov, J. Guo, Y. Z. Zhou, P. W. Gao, D. C. Gu, C. Zhang, S. Jiang, K. Yang, A. Li, Y. C. Li, X. D. Li, X. Dai, Z. Fang, R. J. Cava, L. L. Sun, and Z. X. Zhao, *Breakdown of Three-dimensional Dirac Semimetal State in pressurized $Cd_3As_2$*, arXiv:1410.3213.

[14] K. S. Novoselov, A. K. Geim, S. V. Morozov, D. Jiang, M. I. Katsnelson, I. V. Grigorieva, S. V.

Dubonos, and A. A. Firsov, *Two-dimensional gas of massless Dirac fermions in graphene*, Nature **438**, 197 (2005).

[15] M. Z. Hasan, and C. L. Kane, *Colloquium: Topological insulators*, Rev. Mod. Phys. **82**, 3045 (2010).

[16] X.-L. Qi, and S.-C. Zhang, *Topological insulators and superconductors*, Rev. Mod. Phys. **83**, 1057 (2011).

[17] H. J. Zhang, C.-X. Liu, X.-L. Qi, X. Dai, Z. Fang, and S.-C. Zhang, *Topological insulators in $Bi_2Se_3$, $Bi_2Te_3$ and $Sb_2Te_3$ with a single Dirac cone on the surface*. Nat. Phys. **5**, 438 (2009).

[18] Y. L. Chen, J. G. Analytis, J.-H. Chu, Z. K. Liu, S.-K. Mo, X. L. Qi, H. J. Zhang, D. H. Lu, X. Dai, Z. Fang, S. C. Zhang, I. R. Fisher, Z. Hussain, and Z.-X. Shen, *Experimental Realization of a Three-Dimensional Topological Insulator, $Bi_2Te_3$*, Science **325**, 178 (2009).

[19] Y. F. Zhao, H. W. Liu, X. Guo, Y. Jiang, Y. Sun, H. C. Wang, Y. Wang, H.-D. Li, M.-H. Xie, X.-C. Xie, and J. Wang, *Crossover from 3D to 2D Quantum Transport in $Bi_2Se_3/In_2Se_3$ Superlattices*, Nano Lett. **14**, 5244 (2014).

[20] B.-J. Yang, and N. Nagaosa, *Classification of stable three-dimensional Dirac semimetals with nontrivial topology*, arXiv:1404.0754.

[21] H. M. Yi, Z. J. Wang, C. Y. Chen, Y. G. Shi, Y. Feng, A. J. Liang, Z. J. Xie, S. L. He, J. F. He, Y. Y. Peng, X. Liu, Y. Liu, L. Zhao, G. D. Liu, X. L. Dong, J. Zhang, M. Nakatake, M. Arita, K. Shimada, H. Namatame, M. Taniguchi, Z. Y. Xu, C. T. Chen, X. Dai, Z. Fang, and X. J. Zhou, *Evidence of Topological Surface State in Three-Dimensional Dirac Semimetal $Cd_3As_2$*, Sci. Rep. **4**, 6106 (2014).

[22] M. N. Ali, Q. Gibson, S. Jeon, B. B. Zhou, A. Yazdani, and R. J. Cava, *The Crystal and Electronic Structures of $Cd_3As_2$, the Three-Dimensional Electronic Analogue of Graphene*, Inorg. Chem. **53**, 4062 (2014).

[23] See Supplemental Material for more details.

[24] D. X. Qu, Y. S. Hor, J. Xiong, R. J. Cava, and N. P. Ong, *Quantum Oscillations and Hall Anomaly of Surface States in the Topological Insulator $Bi_2Te_3$*, Science **329**, 821 (2010).

[25] H. Murakawa, M. S. Bahramy, M. Tokunaga, Y. Kohama, C. Bell, Y. Kaneko, N. Nagaosa, H. Y. Hwang, and Y. Tokura, *Detection of Berry's Phase in a Bulk Rashba Semiconductor*, Science **342**, 1490 (2013).

[26] C. Bell, M. S. Bahramy, H. Murakawa, J. G. Checkelsky, R. Arita, Y. Kaneko, Y. Onose, M. Tokunaga, Y. Kohama, N. Nagaosa, Y. Tokura, and H. Y. Hwang, *Shubnikov–de Haas oscillations in the bulk Rashba semiconductor BiTeI*. Phys. Rev. B **87**, 081109 (2013).

[27] M. Novak, S. Sasaki,, K. Segawa, and Y. Ando, *Large linear magnetoresistance in the Dirac semimetal TlBiSSe*, arXiv:1408.2183.

[28] A. A. Abrikosov, *Fundamentals of the Theory of Metals*. (Amsterdam, North-Holland 1988), Vol. **1**.

[29] A. A. Abrikosov, Quantum magnetoresistance, Phys. Rev. B **58**, 2788 (1988).

[30] P. E. Ashby, and J. P. Carbotte, *Theory of magnetic oscillations in Weyl semimetals,* Eur. Phy. J. B, **87**, 1 (2014).

FIG. 1

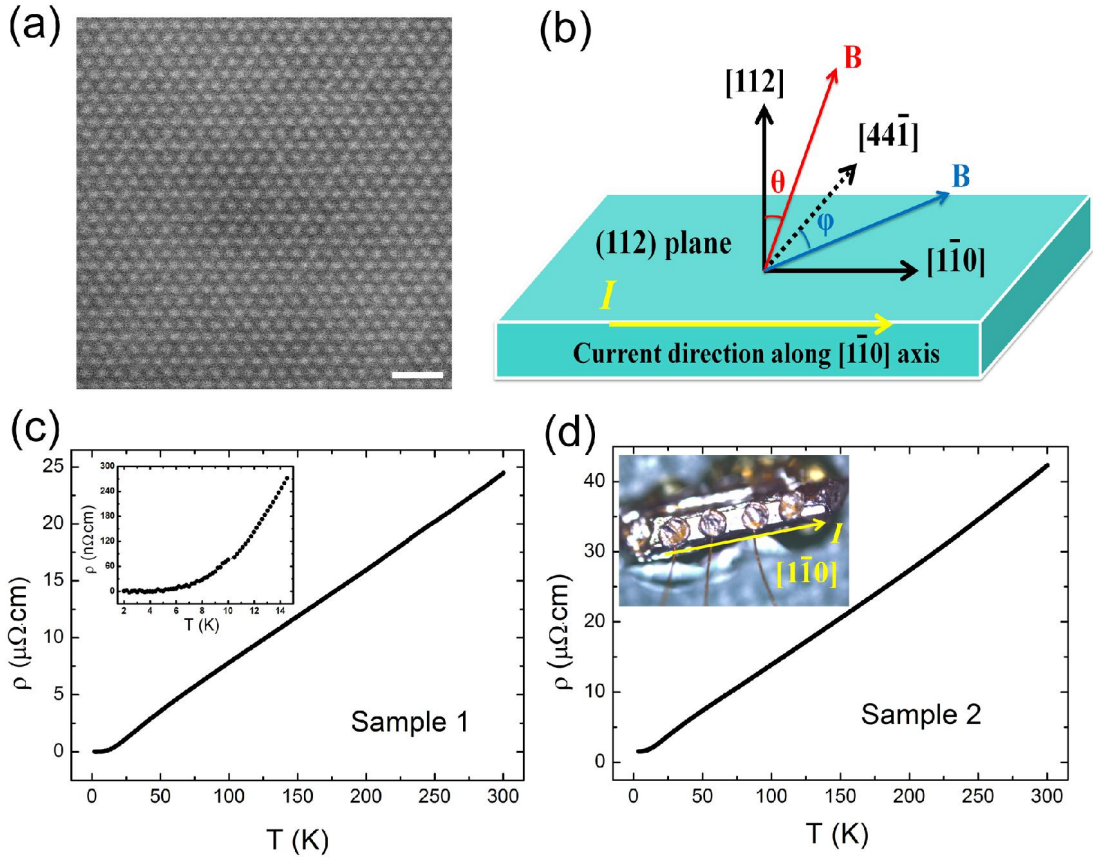

FIG. 2

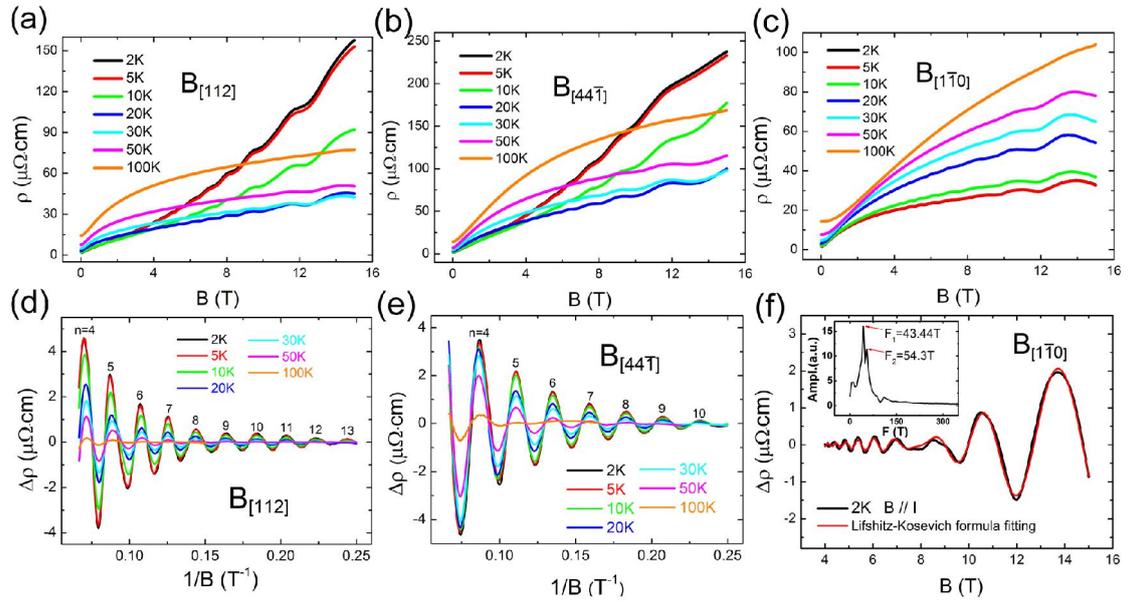

FIG. 3

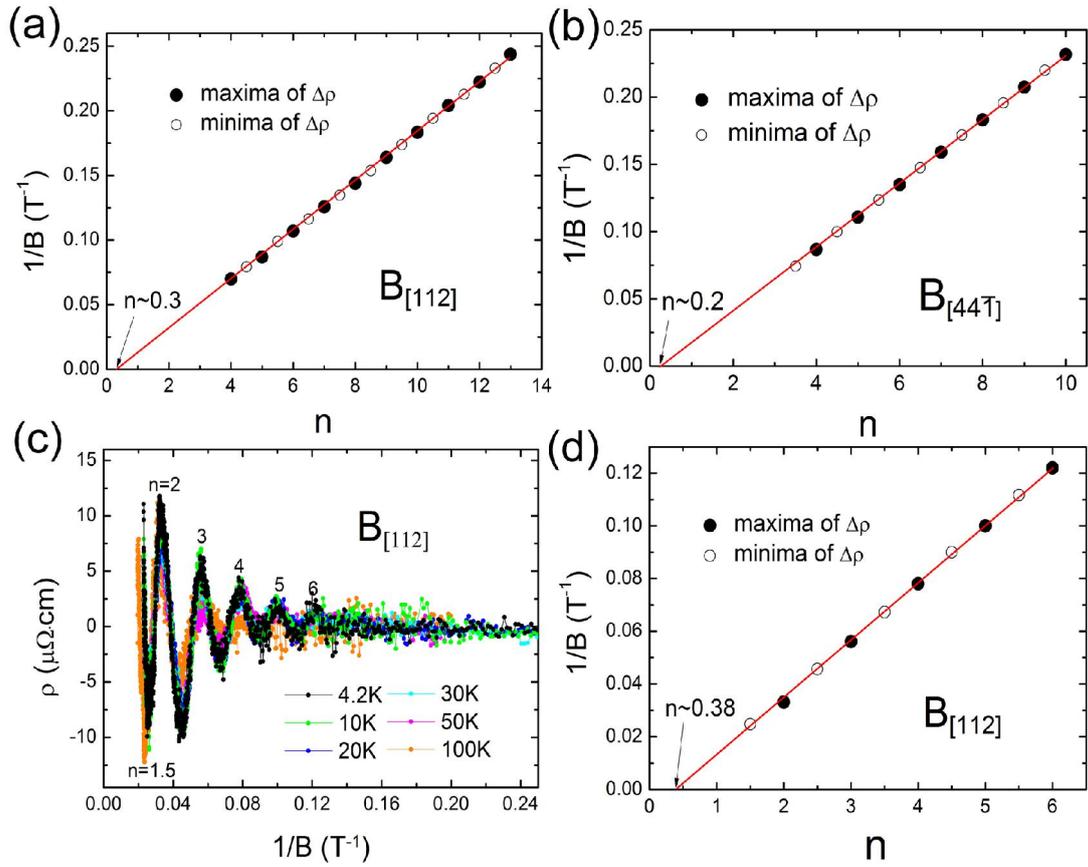

FIG. 4

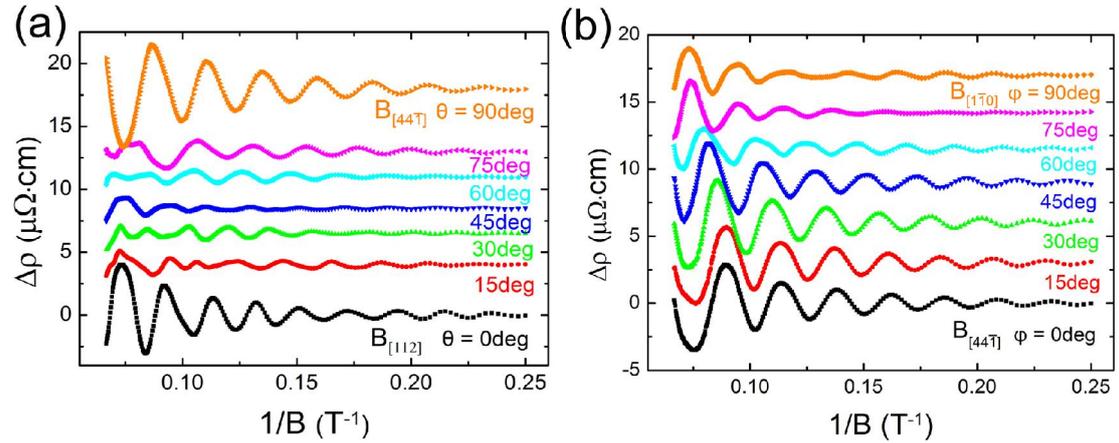

FIG. 5

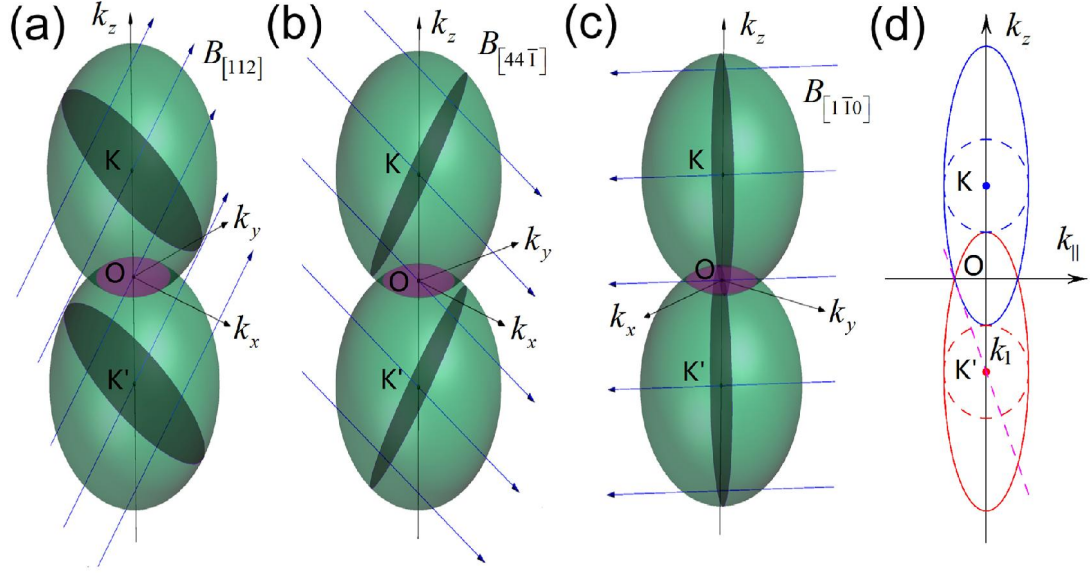

FIG. 6

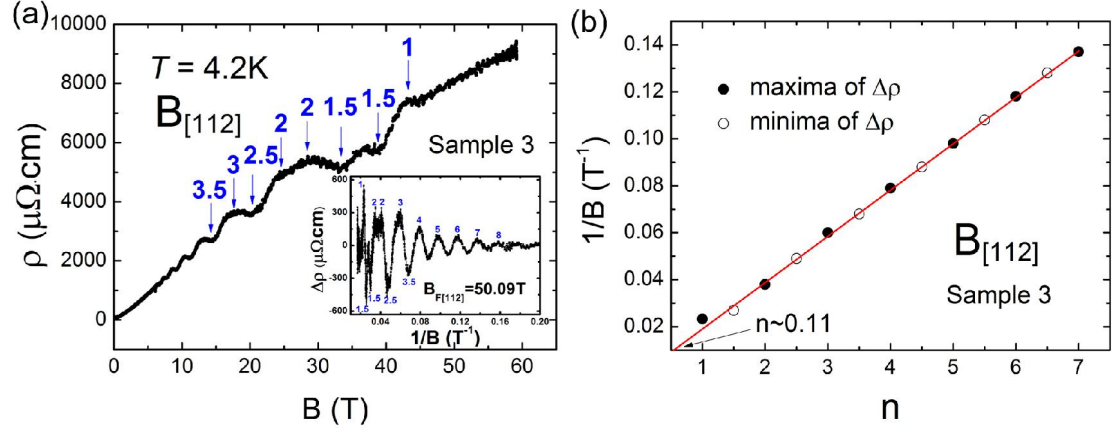

TABLE I:

TABLE I: FFT analysis of SdH oscillations in various magnetic field directions.

| Rotation angle | 0 deg | 15 deg | 30 deg | 45 deg | 60 deg | 75 deg | 90 deg |
|---|---|---|---|---|---|---|---|
| $B_{[112]}$ rotate | 54.30 T | 43.44 T | 48.87 T | 54.30 T | 48.87 T | 43.44 T | 43.44 T |
| to $B_{[44\bar{1}]}$ | – | 54.3 T | 59.73 T | – | – | – | – |
| $B_{[44\bar{1}]}$ rotate | 43.44 T | 43.44 T | 43.44 T | 43.44 T | 43.44 T | 48.87 T | 43.44 T |
| to $B_{[1\bar{1}0]}$ | – | – | – | – | 48.87 T | 43.44 T | 54.30 T |